# Influence of temperature and wavelength on the switchable photovoltaic response of a $BiFe_{0.95}Mn_{0.05}O_3$ thin film

Said Yousfi, Houssny Bouyanfif and Mimoun El Marssi

Laboratoire de Physique de la Matière Condensée EA2081,

Université de Picardie Jules Verne,

33 Rue Saint Leu, 80000 Amiens, France

## Abstract

Photovoltaic (PV) response of epitaxial $BiFe_{0.95}Mn_{0.05}O_3$ thin film grown by pulsed laser deposition has been investigated on a broad range of temperature. Wavelength dependent photovoltaic effect shows the contribution of in gap level states most likely connected to the manganese doping on the B-site of the perovskite unit cells and presence of defects (Bi and O vacancies). The temperature dependent response of the PV response rules out electromigration and/or Schottky barriers as dominant mechanisms. This is corroborated with the observed switchable photovoltaic effect that can be explained either by the depolarizing field or bulk photovoltaic effect. In addition the PV response shows strong correlation with the low temperature polaronic like conduction mechanism and high open circuit voltage (2.5V) is detected in the investigated vertical capacitive geometry.

## Introduction

During the last years, multiferroic materials have gained great attention due to their fundamental physics and possible integration in advanced application like memory device based on the magnetoelectric effect [1]. $BiFeO_3$ (BFO) appears actually as one of the most interesting multiferroic material, due to the room temperature coexistence of ferroelectricity, antiferromagnetism and ferroelasticity. BFO properties are strongly sensitive to imposed strain in thin film form and super-tetragonal phase with giant polarization near the 150μC/cm$^2$ has been observed for instance on appropriate substrates such as $LaAlO_3$ and $YAlO_3$ due to the large compressive epitaxial mismatch [1]. Recently a peculiar photovoltaic effect has also been revealed in BFO with a large open circuit voltage $V_{oc}$ above the band gap [2]. This anomalous PV effect has been studied on different ferroelectric systems and several interpretations have been proposed. The first explanation based on the bulk non-centrosymetry was put forward by Chynoweth in 1956 and applied to $BaTiO_3$, $LiNbO_3$ and $Pb(Zr,Ti)O_3$ [3-5]. Recently another interpretation was applied to BFO and is centred on the role of the domain wall and the analogy between the domain wall (DW) and the pn junction of classical photovoltaic cell [2]. Such domain wall based PV effect in BFO was ruled out by another report and instead fully explain the observed PV response by the bulk R3c non centrosymmetry [6-7]. Reports of the contribution of defects and the influence of the conduction mechanism highlight the complexity of the subject [8-9]. To investigate and better understand the PV effect in BFO we have grown a capacitive structure with BFO deposited on a (001) oriented $LaAlO_3$ substrate buffered with a $SrRuO_3$ bottom electrode. Excellent epitaxial thin films are made on such LAO substrates buffered with SRO bottom electrode and robust ferroelectric properties have been measured on the investigated heterostructure (Remnant polarization close to 70μC/cm$^2$). Rhombohedral BFO film thick enough for good optical absorption and PV response was also chosen (thickness=280nm). The studied heterostructure presents low leakage currents and the conduction mechanism has been elucidated with a change of regime from a nearest neighbour hopping (NNH) above 270K to a variable range hopping mechanism (VRH) below 270K [10]. A full understanding of the charge transport is a prerequisite for deciphering the origin of the PV effect that is reported in this article.

## Experimental details

The epitaxial BFO film was grown on single crystal $LaAlO_3$ $(001)_{pc}$ (LAO) by pulsed laser deposition using an excimer KrF laser (wavelength 248nm) at 6Hz pulse frequency and a fluence of about 1.55J/cm$^2$. The film was deposited at 725°C and 0.05 mbar oxygen pressure. $Bi_{1.1}Fe_{0.95}Mn_{0.05}O_3$ target was used with 10% excess of Bi to prevent Bismuth loss and vacancies during the growth. Target with 5% doping of Mn on the B site has been shown to drastically reduce the leakage current in thin film by lowering the role of oxygen vacancies. $SrRuO_3$ (SRO) 20nm thick bottom electrode was deposited at 650°C and 0.3mbar oxygen pressure. The 100nm thick Pt circular top electrodes of 0.1mm diameter were deposited using a mask by pulsed laser deposition at room temperature and in vacuum ($10^{-6}$ mbar base pressure). Epitaxial growth was obtained and details of the structural characterization can be found elsewhere [10]. The polarization was also studied both by a Sawyer-Tower home-made system and an aixacct system [10]. The I(V) curves were collected using a Keithley 2635 electrometer. An Argon-Krypton tuneable laser was used to illuminate the samples for the PV measurement (457nm to 647nm).

## Results and discussions

The PV effect has been investigated using I(V) curves collected in the dark and under illumination. The voltage was applied to the SRO bottom electrode and swept from +1V to -1V (no differences were found on applying the voltage on the Pt top electrodes). Figure 1 (a) displays the I(V) curve obtained in dark and under laser illumination with different laser power from 5-70mW at 488nm wavelength. The I(V) curve in dark crosses the origin in contrast to the I(V) curves under illumination. An open circuit voltage $V_{oc}$ and short circuit current $J_{sc}$ is observed under illumination with a strong dependence with the laser power. Figure 1 (b) shows the open-circuit voltage $V_{oc}$ and the short circuit current $J_{sc}$ evolution with the laser power. The $V_{oc}$ increases with the laser power up to 50mW in contrast to the monotonic increase of $J_{sc}$ with the laser power. The small decrease of $V_{oc}$ above 50mW can probably be explained by the local overheating.

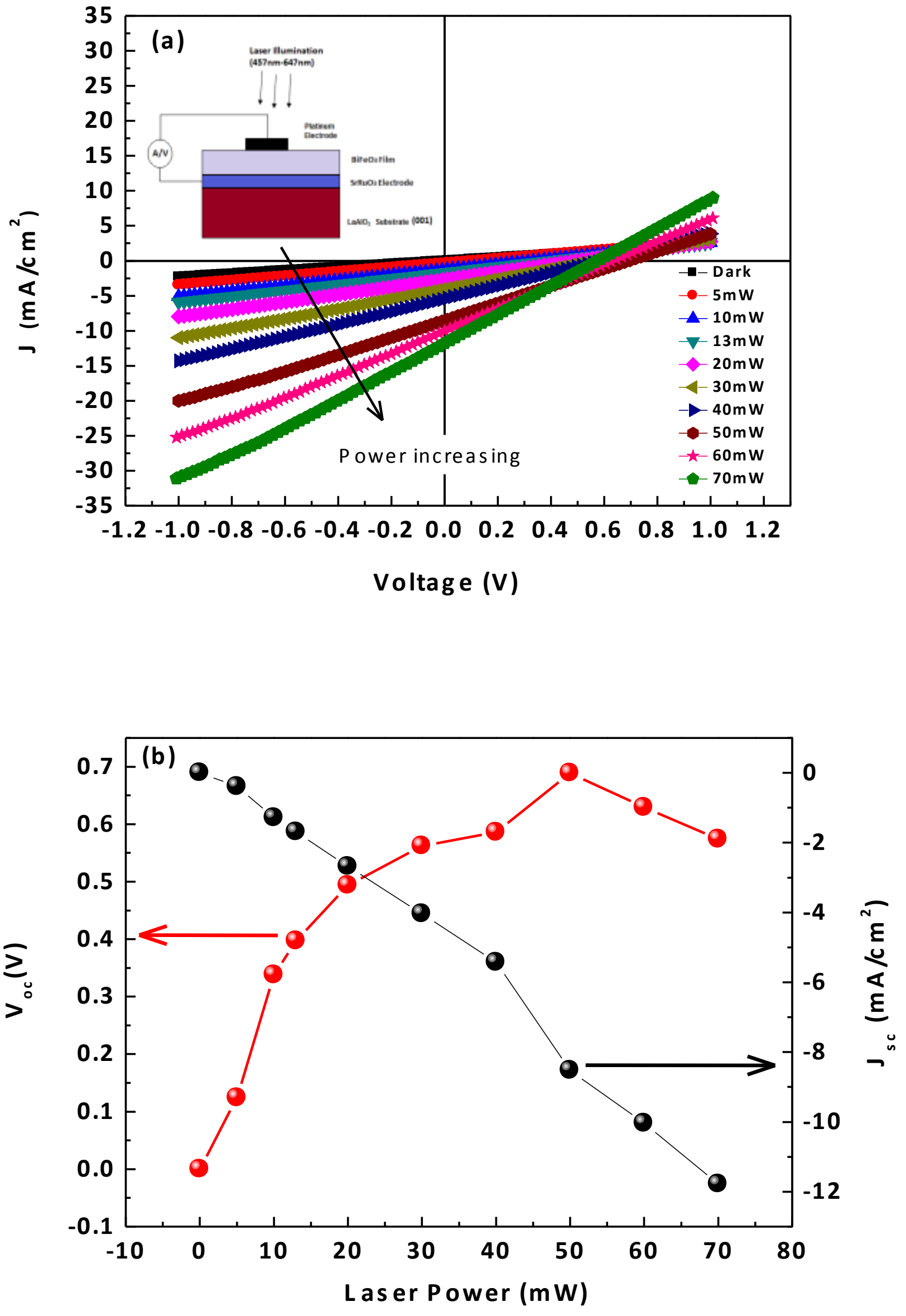

*Figure 1. (a) I(V) curve under laser illumination (488nm) (b) Evolution of the $V_{oc}$ and $J_{sc}$ with the laser power at 488nm wavelength. Insert in (a) shows the diagram of the structure (including substrate, bottom electrode, $BiFe_{0.95}Mn_{0.05}O_3$ film and top electrode).*

The I-V characteristic was measured under laser illumination for different wavelength (457, 488, 514, 568 and 647nm) at a fixed laser power of 13mW to investigate if the PV effect is influenced by the wavelength of light. The results are reported in Figure 2 (a) and the evolution of $V_{oc}$ and $J_{sc}$ with wavelength are presented in Figure 2 (b) showing a photovoltaic effect whatever the wavelength. An increase of $V_{oc}$ and $J_{sc}$ is observed with decreasing the wavelength.

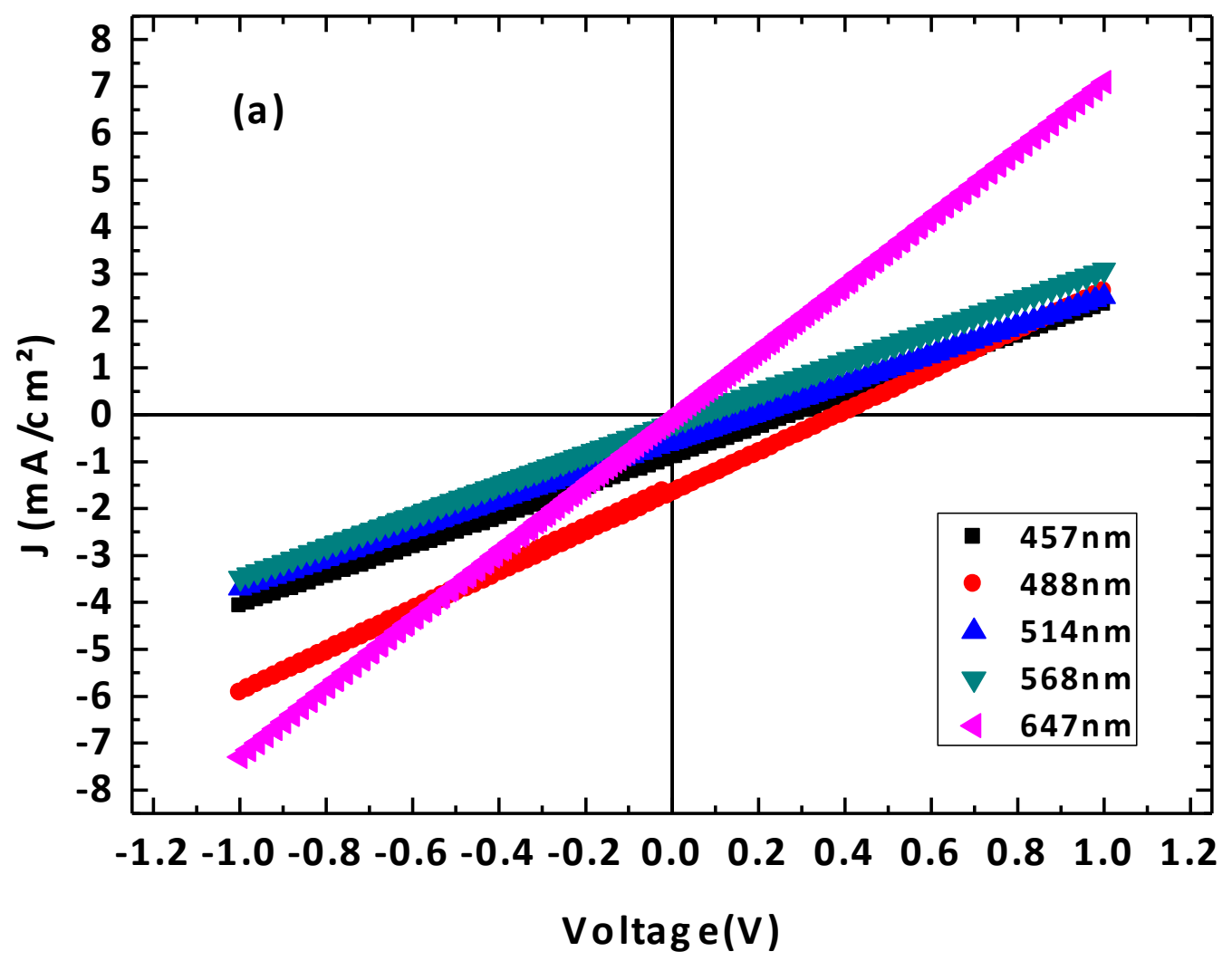

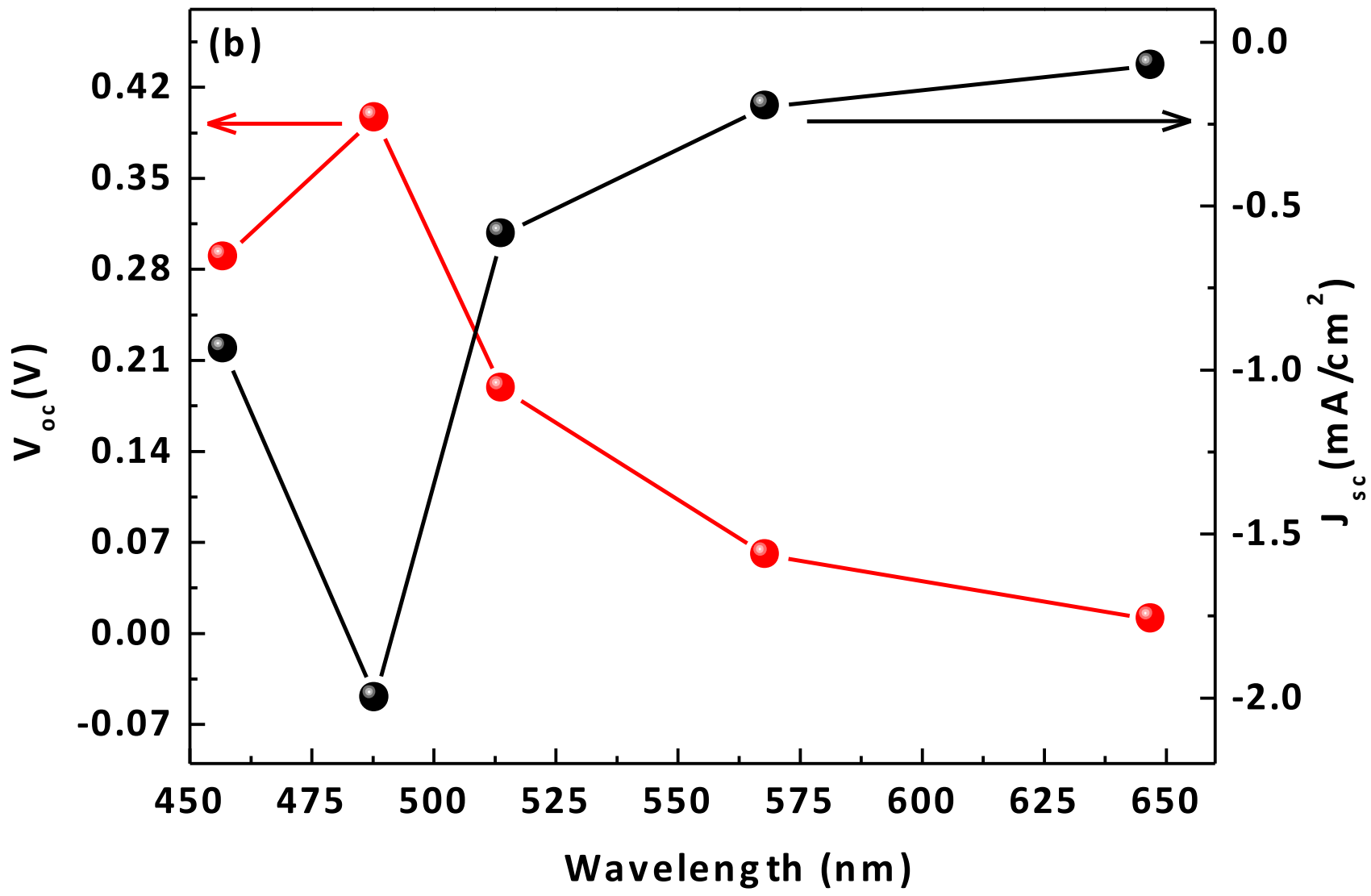

*Figure 2. (a) I(V) curves measured under different wavelength (13mW). (b) Evolution of the $V_{oc}$ and $J_{sc}$ with the wavelength.*

The existence of a photovoltaic effect for all the laser wavelength from 457nm (2,71ev) to 647nm (1,92ev) is in contradiction with the fact that BFO exhibits direct gap energy near 2,7ev [2][11]. The presence of a distribution of defect trap levels in the band gap is responsible for this photovoltaic effect and oxygen vacancies and the 5% Manganese doping are obviously candidates for inducing in gap level states [12] [13]. As shown in Figure 2(b) a decrease of Voc and Jsc by decreasing the wavelength from 488 nm to 457 nm is observed and may be connected to sub-band levels just as observed by Akash Bhatnagar *et al.* [14]. It is not clear yet whether these sub-band levels correspond to oxygen vacancies and/or manganese doping.

The detected PV effect can be caused by the inhomogeneous potential at the film/electrode interfaces, the depolarizing field effect or the PV bulk non-centrosymmetric mechanism. Electromigration of mobile vacancies are also impacting the I(V) characteristics under illumination and it is difficult to disentangle the many process at work. In order to better understand the PV effect we investigated the effect of short applied voltage pulses (width of 90ms) of different amplitude and sign. The I(V) curves obtained under illumination after applying positive/negative pulses of increasing amplitude have been measured and the PV characteristics ($V_{oc}$ and $J_{sc}$) are shown in Figure 3.

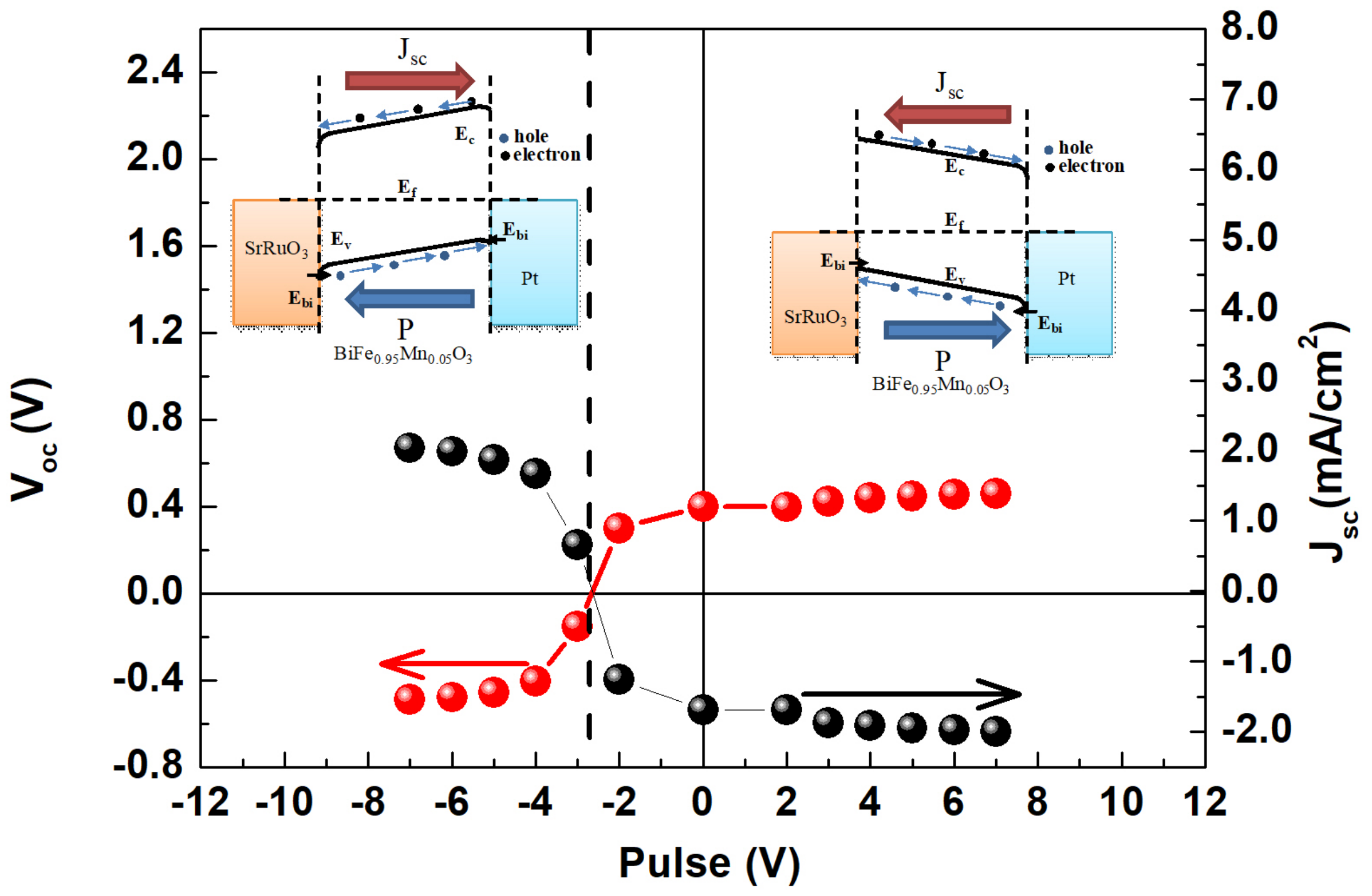

*Figure 3. Evolution of $V_{oc}$ (red spheres) and the $J_{sc}$ (black spheres) with applied positive and negative pulses of various amplitudes (illumination @488nm and 13mW) applied on SRO bottom electrode. Vertical dashed line highlights the origin from which absolute voltages are estimated to*

*switch the PV effect. This vertical dashed line also delineates the region of up and down polarizations and simplified schematics of the band structure, direction of polarization and Jsc are provided in top right and left inserts ($E_f$: Fermi level, $E_{bi}$: built in field, $E_c$: conduction band, $E_v$: valence band).*

A switchable PV effect is clearly observed. After large applied positive pulses $V_{oc}$ and $J_{sc}$ saturate at 0.48V and -2mA/cm$^2$ respectively. Opposite values are obtained after applying sufficient negative pulses. An interesting observation is the spontaneous PV effect already observed in Figure 1 and corresponding to the 0V pulse on the Figure 3 at which negative $J_{sc}$ and positive $V_{oc}$ are measured. A negative shift of -2.66V (vertical dashed line) of the whole pattern is evidenced suggesting the presence of in-built electric field (relative to the SRO bottom electrode) in the as-grown film. Such in-built field has been observed in the P-E loop with an imprint with similar sign. Self polarization with upward ferroelectric polarization was indeed observed on the thin film. Self polarization can be caused by the presence of oxygen vacancies, inhomogeneous strain and different bottom and top electrodes and we should precise that the coercive voltage for the investigated film is about 3.5V. This is in good agreement with the absolute voltage relative to the -2.66V needed to fully switch the PV response. Indeed whatever the orientation, the PV response is fully saturated only after applying pulses of nearly 3.5V relative to the vertical dashed line on Figure 3. Following Zheng *et al.* it is possible to estimate the relative contribution of the depolarizing field ($J_{dep}=(J_{up}-J_{down})/2$) from a possible in built bias ($J_{Bias}=(J_{up}-J_{down})/2$) with $J_{up(down)}$ the short circuit current for up(down) polarization [15]. The estimated values are $J_{dep}$=-2mA/cm² and $J_{Bias}$~0mA/cm² showing a ferroelectric or bulk-like origin of the PV effect. In fact this estimation does not however distinguish between the pure non centrosymmetry and the depolarizing field contribution. A simplified schematic diagram of the band structure is provided on figure 3 (top right and left inserts). Considering the high resistance of the thin film $E_f$ is close to the middle of the gap (BFO p type [11]). As discussed below and elsewhere SRO/BFO and Pt/BFO Schottky barriers are not controlling the transport in the heterostructure and are deliberately made small on figure 3 to emphasize the primary role of the polarization on the PV response ($E_{bi}$: Schottky barriers on figure 3; see [10] for more details on conduction mechanism). Although it is therefore natural to attribute such switchable PV effect to the switching of the spontaneous polarization it is not possible to rule out another contribution and to further understand the origin of the PV response we have explored the influence of temperature on I(V) measurements. Figure 4 shows the $V_{oc}$ and $J_{sc}$ extracted from the I(V) curves under illumination for temperatures between 93 to 393K. The I(V) curves were collected at the same power of 13mW and at the five  wavelengths 457, 488, 514, 568 and 647nm.

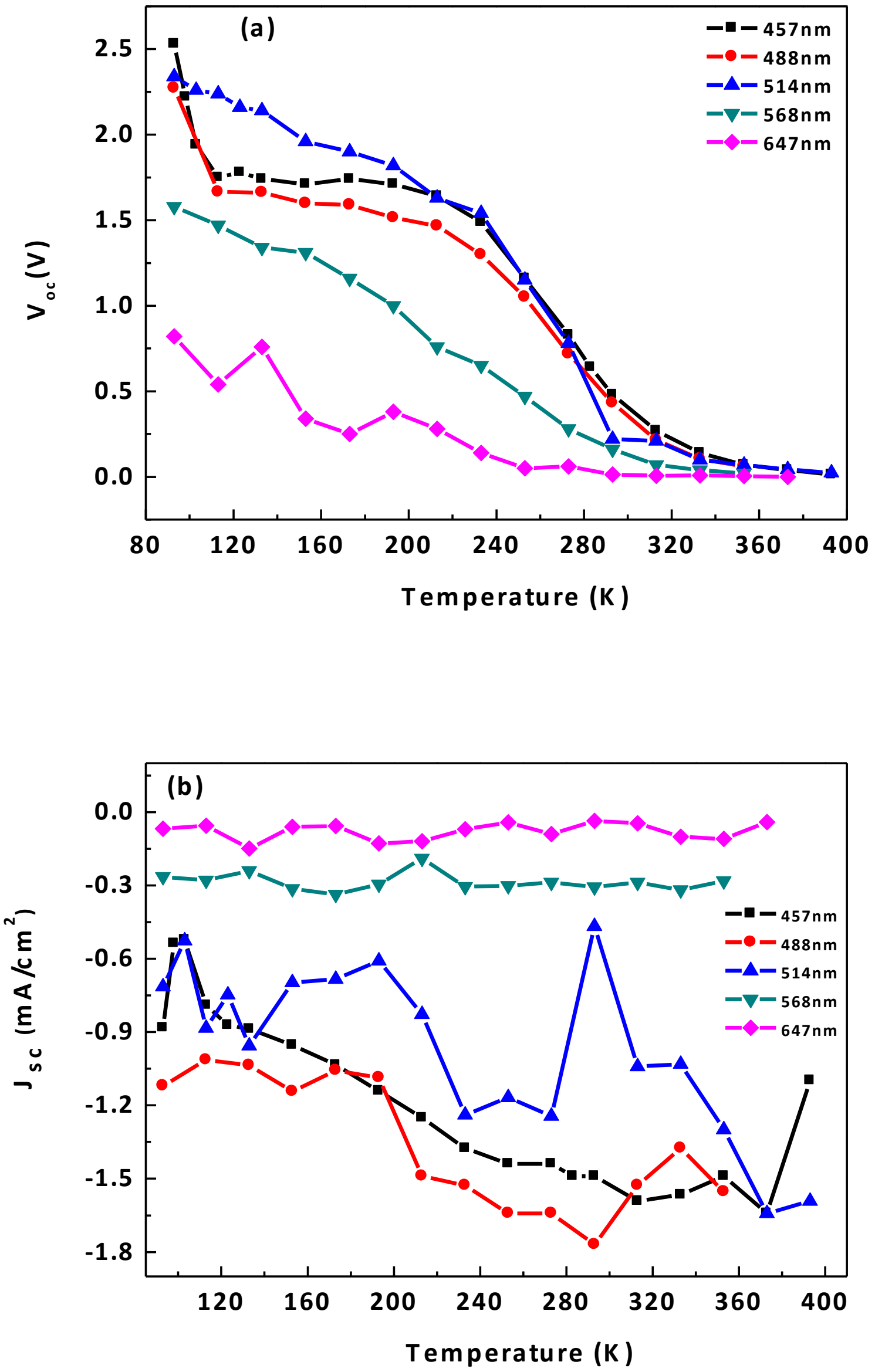

*Figure 4. Temperature dependence of $V_{oc}$ (a) and $J_{sc}$ (b) at five wavelengths 457, 488, 514, 568 and 647nm.*

An increase of $V_{oc}$ is observed on cooling down. The increase is particularly clear for the 457nm, 488nm and 514nm wavelengths. The smaller increase of Voc at 568 nm and 647 nm may be due to the lower PV effect at such wavelength of light. Concerning the Jsc, no changes are detected for the 568 nm and 647 nm probably due to the very low PV response at such wavelength. For the 457nm, 488nm and 514nm wavelengths Jsc shows a decrease on cooling. Such decrease may be due to the increase of the bandgap on cooling and the subsequent decrease of photo-induced transition electron-hole pair [16]. The increase of $V_{oc}$ on cooling is opposite to the PV behaviour revealed by Ge *et al.* [17]. These authors evidenced an increase of $V_{oc}$ with temperature connected with electromigration of oxygen vacancies. Our results strongly suggest that electromigration of oxygen vacancies are not the driving force of the PV effect but are not in contradiction with conclusion made by Ge *et al.*. Indeed the observations made by Ge *et al.* [17] and in this work indicate that the PV mechanism strongly depends on the stoechiometry and nature of the sample (thin film, crystal, ceramics show a diversity of PV response [18]). Note that we measured a $V_{oc}$ above 2.5V at the lowest temperature (@ 488nm wavelength) suggesting that the Schottky barriers at the film electrode interfaces may not be the dominant mechanism causing PV response. Schottky barriers are whatever the conduction mechanism present and barriers of about 0.5-0.6eV are expected for SRO/BFO and Pt/BFO interfaces (See [9] on which the following discussion is based). The two barriers are similar to two back-to-back diodes and only one blocking barrier remains when effect of polarization charges is taken into account. Depending on the orientation of the ferroelectric polarization, the SRO/BFO or Pt/BFO Schottky barrier and local electric field may induce PV effect whose amplitude of open circuit voltage cannot exceed the bandgap value [9]. The observed $V_{oc}$ of 2.5V is probably not arising from a Schottky mechanism and the modification of the Schottky barrier by the polarization. These results are in agreement with the transport mechanism investigated on the same film for which bulk limited conduction mechanisms were revealed [10]. Variable range hopping (VRH) is observed below 250-270K while nearest neighbour hopping (NNH) is detected above 250-270K and interestingly the $V_{oc}$ values show a change at similar temperatures on Figure 4. The electrical transport is not controlled by the Schottky barriers at the interfaces for this thin film and a change of regime for the hopping mechanism at 250-270K has been evidenced. This change of regime probably explains the $V_{oc}$ behaviour detected on Fig. 4 in agreement with equation (1) given below by Fridkin [19].

$$V_{oc} = J_{sc}\left(\frac{1}{\sigma_d+\sigma_{ph}}\right)L \qquad (1)$$

With $\sigma_d$: the dark conductivity and $\sigma_{ph}$: the photoconductivity

While the conduction mechanism is involved in the $V_{oc}$ behaviour with temperature and gives clues on the most likely uniform distribution of the applied voltage drop in the film it is not obvious yet what is the driving mechanism of the PV effect. Depolarizing field effect and bulk photovoltaic effect relative contributions are indeed difficult to separate in such geometry and these two different mechanisms may be simultaneously at work.

**Conclusion**

A switchable PV effect is detected in epitaxial (001) oriented BFO 280nm thick film of rhombohedral structure. Wavelength dependent PV response strongly suggests in gap defect states contribution. Large $V_{oc}$ at low temperature from temperature dependent measurements allows ruling out Schottky barriers and/or electromigration of oxygen vacancies as main driving forces to the PV activity. Depolarizing field effect and bulk photovoltaic effect may be simultaneously at work and cannot be distinguished in the investigated geometry. Interestingly polaronic like conduction mechanism are correlated with the $V_{oc}$ variation with temperature and full understanding of the photo-response of ferroelectrics (photostriction and photovoltaic effects) may involved such connection between the photo excited species, nature of charge carriers and the local structural inversion asymmetry.

**Acknowledgements:** This work has been funded by the region of Picardy (project ZOOM), the French government (MESR) and the European commission (H2020 RISE "ENGIMA n° 778072").